\RequirePackage[2020-02-02]{latexrelease}
\documentclass{iau}

\usepackage{amsmath}
\usepackage{graphicx}
\usepackage{multirow}

\lefttitle{S. Torniamenti}
\righttitle{A novel generative method for star clusters}

\jnlPage{1}{7}
\jnlDoiYr{2021}
\doival{10.1017/xxxxx}

\aopheadtitle{Proceedings IAU Symposium}
\editors{C. Sterken,  J. Hearnshaw \&  D. Valls-Gabaud, eds.}

\title {A novel generative method for star clusters \\ from hydro-dynamical simulations}

\author{Stefano Torniamenti$^{1,2,3}$}  

\affiliation{$^{1}$ Physics and Astronomy Department Galileo Galilei, University of Padova, Vicolo dell'Osservatorio 3, I--35122, Padova, Italy
\\
$^{2}$ INFN- Sezione di Padova, Via Marzolo 8, I--35131 Padova, Italy 
\\
$^{3}$ INAF, Osservatorio Astronomico di Padova, vicolo dell'Osservatorio 5, 35122 Padova, Italy
\\
email: \email{stefano.torniamenti@studenti.unipd.it}}
\setcitestyle{numbers}

\begin{document}

\begin{abstract}
Most stars form in clumpy and sub-structured clusters. These properties also emerge in hydro-dynamical simulations of star-forming clouds, which provide a way to generate realistic initial conditions for $N-$body runs of young stellar clusters. However, producing large sets of initial conditions by hydro-dynamical simulations is prohibitively expensive in terms of computational time.
We introduce a novel technique for generating new initial conditions from a given sample of hydro-dynamical simulations, at a tiny computational cost. In particular, we apply a hierarchical clustering algorithm to learn a tree representation of the spatial and kinematic relations between stars, where the leaves represent the single stars and the nodes describe the structure of the cluster at larger and larger scales. This procedure can be used as a basis for the random generation of new sets of stars, by simply modifying the global structure of the stellar cluster, while leaving the small-scale properties unaltered.

\end{abstract}


\maketitle

\section{Introduction}
A large fraction of star formation happens in clusters or associations \cite{ladalada2003}. These star-forming systems are characterized by complex phase-space distributions, where sub-structures \cite{larson1996}, fractality \cite{Cartwright09,Kuhn19}, relative sub-clump motions \cite{CantatGaudin19}, expansion (due to gas expulsion, \citealp{hills80}),
and possibly rotation \cite{henaultbrunet12} are observed. 
The early evolution of these stellar systems is of fundamental importance for the comprehension of the present-day structure of older open and globular clusters, and can be modeled in a realistic way by means of direct $N-$body simulations. However, adequate initial conditions, able to take into account the observed phase-space complexity, are necessary for a correct comprehension of this evolutionary phase.
In this sense, spherical equilibrium models like Plummer spheres \cite{plummer} or King models \cite{king} cannot represent the initial spatial and kinematic distributions of young star clusters. A natural way to reproduce the observed properties of these stellar systems is by means of hydro-dynamical simulations of collapsing molecular clouds, where they naturally emerge \cite{ballone2020,ballone2021,bate09,klessen12,wall19}. However, running  hydro-dynamical simulations including all the relevant physics is computationally very expensive, and producing large sets of initial conditions would turn to be prohibitive in terms of computational time.
 
Here, we introduce a novel method for producing new initial conditions for $N-$body runs without running additional independent hydro-dynamical simulations. Our approach relies on applying a clustering algorithm to an existing set of initial conditions. In particular, we adopt a hierarchical clustering algorithm, which
learns a tree-like representation of the stellar cluster, describing its structure at different scales. In our generative model, new stellar clusters are then obtained by modifying selected nodes of the hierarchical tree.

\section{Sink particle distributions} \label{sec_sink}

We consider the sink particle distributions (also named as stars in the following) from $10$ smoothed-particle hydro-dynamical simulations of molecular clouds, performed by Ballone {\it et al.} (2020) \cite{ballone2020}.
These simulations are initialized as spherical molecular clouds with total gaseous mass ranging between $10^4 \, \mathrm{M_{\odot}} $ and $10^5 \, \mathrm{M_{\odot}}$, uniform temperature $T=10 \, \mathrm{K}$ and uniform density $\rho = 2.5\times 10^2 \, \mathrm{cm^{-3}}$. Star formation is implemented during the simulation by means of a sink particle algorithm \cite{bate95}.
The star clusters are the result of the instantaneous gas removal at 3 Myr, mimicking the impact of the first supernova explosions (no stellar feedback was included in the simulations). 
We refer to Ballone et al. (2020,2021) \cite{ballone2020,ballone2021} for more details about the hydro-dynamical simulations.
Table \ref{tab_sink} resumes the main properties of the sink particle distributions under consideration.

\begin{table*}[h!]
\centering
{\tablefont\begin{tabular}{@{\extracolsep{\fill}}lllllllllll}
\midrule
 &\texttt{m1e4} & \texttt{m2e4} & \texttt{m3e4} & \texttt{m4e4} & \texttt{m5e4} & \texttt{m6e4} & \texttt{m7e4} & \texttt{m8e4} &
\texttt{m9e4} & \texttt{m1e5} \\ \hline
$N_{\rm s}$  & 2523  & 2571 & 2825 & 2868 & 2231 & 3054 & 4214 & 2945 & 3161 & 3944 \\
$M_{\rm s}$ $ \left[10^3 \, {\rm M}_\odot\right]$ & 
$4.2$ &  $6.7$ & $10.3$ &  $14.4$ & $14.1$ & $20.4$ &  $31.5$ & $28.3$ & $30.5$ & $38.0$ \\
$M_{\rm mc}$ $\left[10^3 \, {\rm M}_\odot\right]$ & $10.0$ &   $20.0$ & $30.0$  & $40.0$  & $50.0$  & $60.0$  & $70.0$  & $80.0$ & $90.0$  & $100.0$ \\
$\epsilon_{\rm sf}$ & $0.42$ & $0.33$ & $0.34$ & $0.36$ & $0.28$ & $0.34$ & $0.45$ & $0.35$ & $0.34$ & $0.38$ \\
\midrule
\end{tabular}} \label{tab_sink}
\caption{\textit{Columns}: [1] number and [2] total mass of the sink particles, [3] mass of the parent molecular cloud, and [4] the resulting star formation efficiency
}\end{table*}

\section{Hierarchical clustering}

Clustering algorithms are a class of unsupervised machine learning methods. In general, clustering identifies similar instances in a given sample and assigns them to groups (or clusters).
In the specific case of hierarchical clustering, the algorithm proceeds in a hierarchical way, by connecting the most similar pair of clusters, starting from the individual instances (in this case the single stars), until a certain number of groups is reached\footnote{When the hierarchy is built from the bottom up, the algorithm is also referred to as agglomerative clustering algorithm.} \cite{kaufman}. To identify similar elements in the sample, hierarchical clustering is provided with a similarity prescription, called linkage. 
For this case, we make use of Ward's linkage, which merges two clusters such that the variance within all clusters increases the least. This often leads to clusters that are relatively equally sized. For more details on such choice, we refer the reader to Torniamenti {\it et al.} (2022) \cite{torniamenti2022}. Also, we use the implementation offered by the {\sc scikit-learn} library \cite{pedregosa}.

\begin{figure*}
\includegraphics[width=\textwidth]{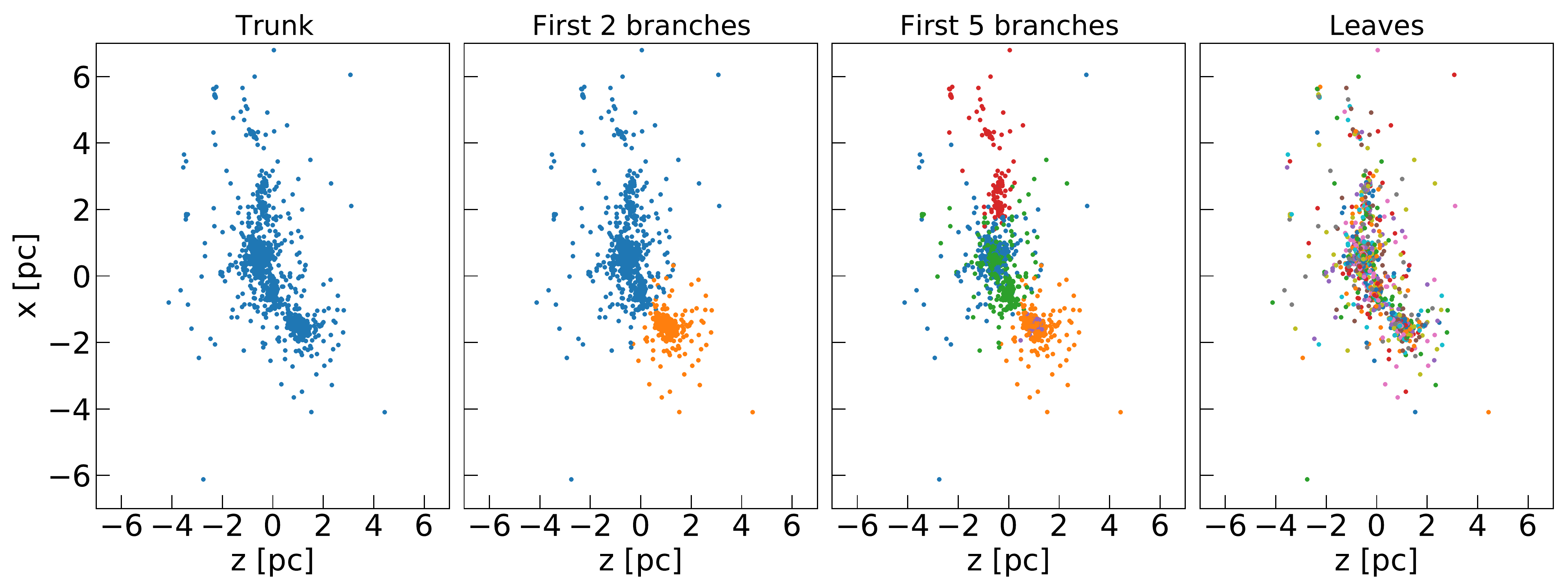}
\caption{Different levels in the hierarchical tree of the {\tt m1e4} simulation. The panels show different levels of the the tree: [1] the trunk, [2] the first two branches, [3] the first five branches, and [4] the leaves.}
\label{fig_m1e4}
\end{figure*}

\subsection{Application to stellar clusters}

We applied hierarchical clustering to the stellar clusters introduced in Section~\ref{sec_sink}. Before applying the algorithm, we scaled the positions and the velocities by their standard deviations. 
Figure \ref{fig_m1e4} shows different levels of the cluster hierarchy for the {\tt m1e4} cluster. The clumps are organized into a hierarchical tree-like structure $\mathcal{T}$, where the trunk contains the whole set of stars and each subsequent node is a two-way split (with each branch being a sub-clump), down to the leaves, representing individual stars.

The hierarchical construction allows to identify groups of similar instances in the distribution of stars as well as  drawing information about the structure of the star system at different scales. To describe the relevant physical properties of the cluster by means of the tree formalism, at each node we evaluate the distance vector between the centres of mass of the clumps $\mathbf{l}_\mathrm{i}$, their relative velocity vector $\mathbf{u}_\mathrm{i}$, and the mass ratio between the two clumps {\bf $q_\mathrm{i}$}, defined as the ratio between the lightest of the two resulting groups and the total mass of the node. 

\begin{figure*}
\includegraphics[width=\textwidth]{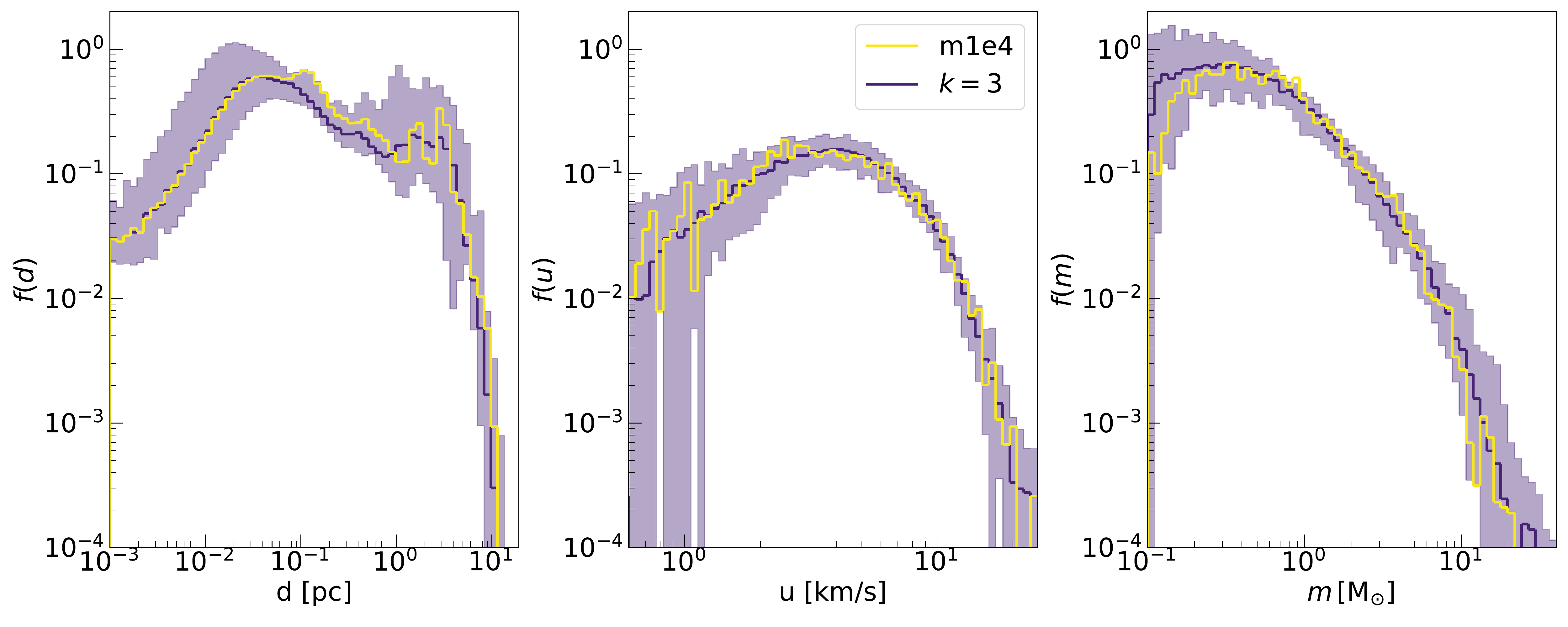}
\caption{Distributions of inter-particle distances $f(d)$ (left), velocities $f(v)$ (center), and masses $f(m)$ (right) for the sink particles taken from the {\tt m1e4} simulation (thick yellow line) and the distributions of new generations obtained by replacing the first 2 nodes (corresponding to $k=3$, purple). The shaded area encloses the distribution of the new generations, and the solid line is the median of the distribution.}
\label{fig_dist}
\end{figure*}

\section{Generative method}
Each node of the tree, $\mathcal{T}_i$, describes the relations between two sub-clumps departing from a parent branch. As a consequence, the relevant quantities $\mathbf{l}_\mathrm{i}$, $\mathbf{u}_\mathrm{i}$, and $q_\mathrm{i}$ can be used as instructions to progressively split clumps of stars in the phase space, starting from one reference mass. 
During this procedure, we can modify some selected nodes of the tree to obtain a new and different realization of the stellar cluster. The way in which the nodes are modified depends on what scales of the systems we want to preserve or alter. In our case, we aim to obtain new macroscopic configurations by preserving the small scale properties (such as their complex fractal structure), which make our clusters so realistic. 
For this reason, to obtain a new realization, we replace the first nodes of a reference tree, which describe the large scale distribution of sub-clumps, with the same quantities drawn from other trees. The method consists in the following steps:
\begin{itemize}
    \item We consider a reference tree $\mathcal{T}$, and we replace the quantities $\mathbf{l}_\mathrm{i}$, $\mathbf{u}_\mathrm{i}$ and $q_\mathrm{i}$, associated to the first $i < k$ nodes, with the same-level quantities from the tree $\mathcal{T^\prime}$, learned from another set of sink particles. 
    Here, we consider $k=3$. For more details on such choice, we refer the reader to Torniamenti {\it et al.} (2022) \cite{torniamenti2022}. 
    
    \item We consider one particle containing the total mass of the cluster we want to generate, placed at rest in the origin of the coordinate system. This particle is split into two particles, such that the resulting mass ratio is $q_{\mathrm{1}}$ (the mass ratio relative to the first node of the new tree). The new positions and velocities are assigned such that their centre of mass is at rest in the origin of the system, and their distance and relative velocity vectors are  $\mathbf{l}_\mathrm{1}$ and $\mathbf{u}_\mathrm{1}$, respectively.
    \item At each step $i$, we split a chosen particle into two new particles with mass ratio $q_\mathrm{i}$, at a distance $\mathbf{l}_\mathrm{i}$ from each other, moving with relative velocity $\mathbf{u}_\mathrm{i}$. The particle-to-split is chosen by considering the same order of splitting as the original reference tree.
    This splitting procedure is then repeated until a cluster with the same number of particles as the reference one is obtained.
    \item Finally, we remove the very low-mass stars (which may result in planet-sized objects) by setting a cutoff mass to the minimum mass of the original stars on which $\mathcal{T}$ was learned. 
\end{itemize}

\begin{figure*}
\includegraphics[width=\textwidth]{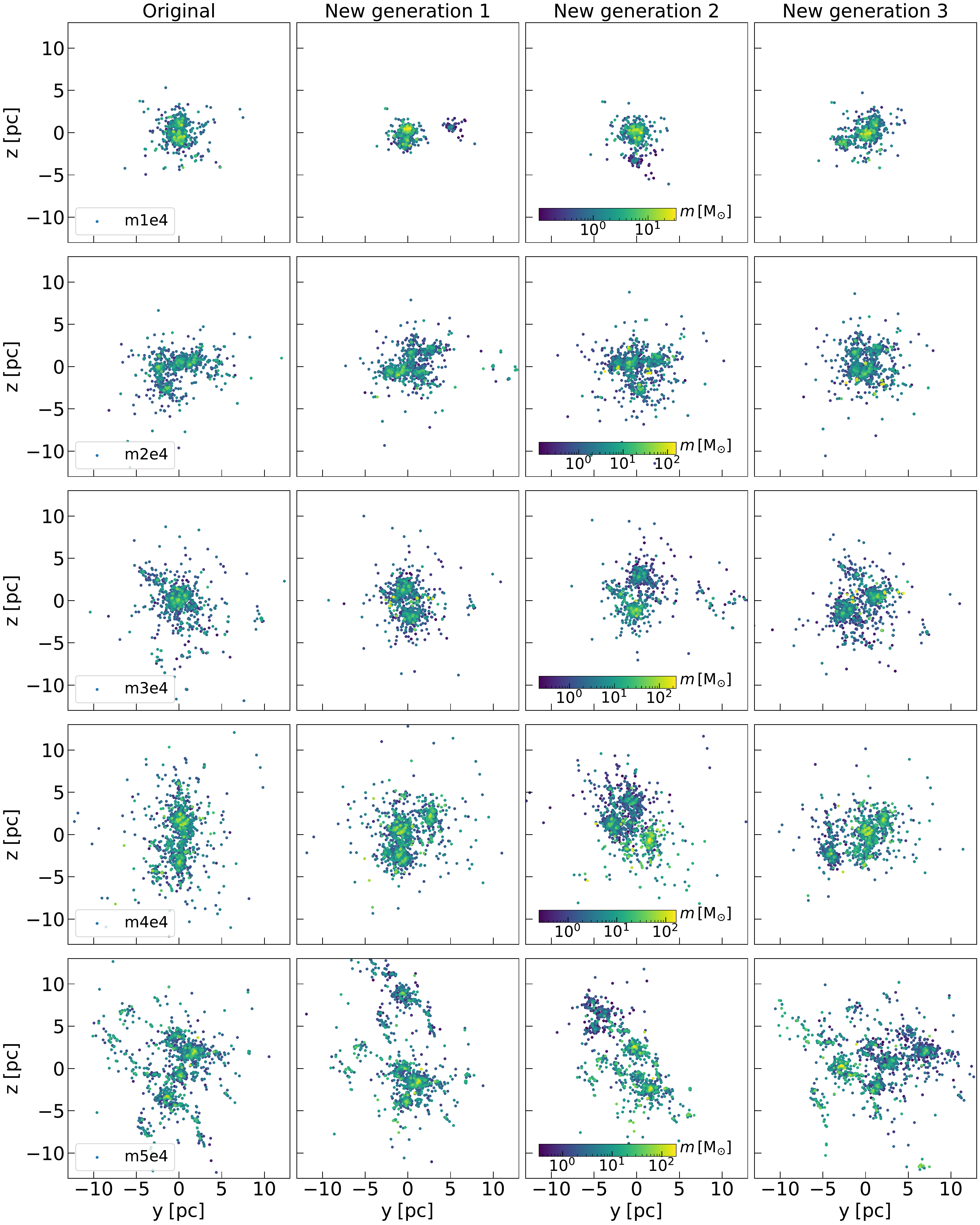}
\caption{Projections in $y-z$ of the 5 least massive star clusters (left), and of three different generated clusters per each. The colour map marks the mass of the individual stars.}
\label{fig_ALL_1}
\end{figure*}

\begin{figure*}
\includegraphics[width=\textwidth]{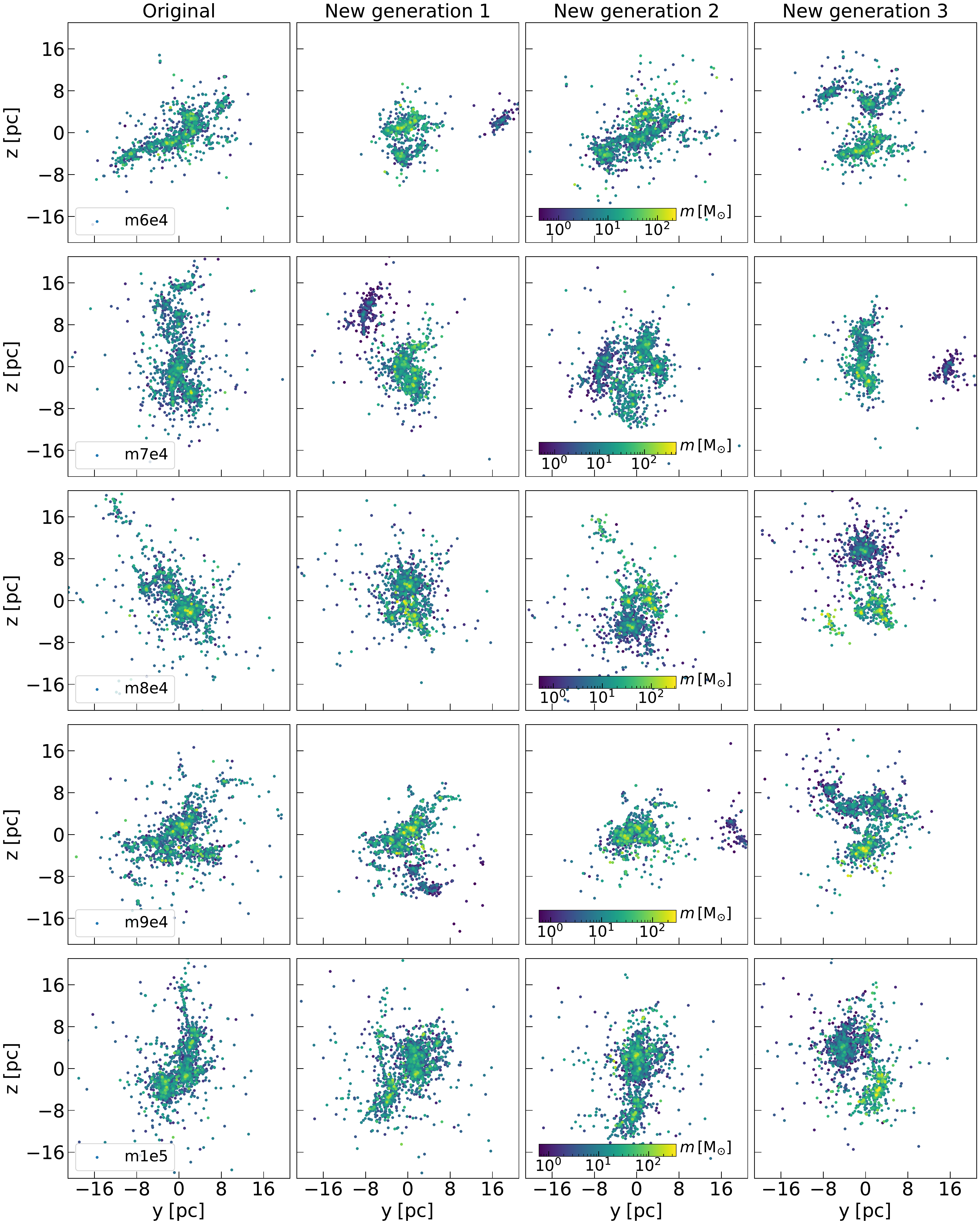}
\caption{Same as Fig. \ref{fig_ALL_1}, but for the 5 most massive star clusters.}
\label{fig_ALL_2}
\end{figure*}

In Fig. \ref{fig_dist} we compare the distance, velocity and mass distributions of {\tt m1e4} to those of a set of new generations. All the distributions of the new realizations are consistent with those of the original simulation.
In particular, the new distance distributions are altered at large scales but, moving toward small distances, they recover  the same trend as the original simulation, as meant for this method. 
Figure \ref{fig_ALL_1} and \ref{fig_ALL_2}  show the  spatial distributions of the original cluster and of three new generations per each, for all the sink particle distributions of our sample. The new generations are qualitatively indistinguishable from the original clusters.

\section{Summary}
We introduced a new method for generating a number of new realizations from a given set of initial conditions from hydro-dynamical simulations. This method is based on a hierarchical clustering algorithm, which learns a tree-like representation of the stellar system. 
This tree encodes the structural properties of the sink particle distributions and can be turned into new macroscopic realizations by modifying its first branches. This procedure results in different large scale realizations (e.g., the number of main clumps and their distances), while approximately preserving the characteristics of the small scale structure responsible for most of dynamical evolution. The new realizations are qualitatively similar to the original simulations when visualized in the three-dimensional space, and present consistent velocity, mass and pairwise distance distributions. 

This generative method leads to a speedup in computation of several orders of magnitude: generating initial conditions from hydro-dynamical simulations, in fact, requires hundreds of thousands core hours per simulation, while our procedure takes about some core seconds to generate a new realization. Also, our scheme is very flexible, allowing to set how deep we alter the tree structure by choosing the number of initial branches we modify.

\section{Acknowledgements}
This project is partially supported by European Research Council for the ERC Consolidator grant DEMOBLACK, under contract no. 770017, and by the European Unions Horizon 2020 research and innovation program under the Marie Skłodowska-Curie grant agreement No.896248.

\end{document}